\def\year{2020}\relax
\documentclass[letterpaper]{article} % DO NOT CHANGE THIS
\usepackage{aaai20}  % DO NOT CHANGE THIS
\usepackage{times}  % DO NOT CHANGE THIS
\usepackage{helvet} % DO NOT CHANGE THIS
\usepackage{courier}  % DO NOT CHANGE THIS
\usepackage[hyphens]{url}  % DO NOT CHANGE THIS
\usepackage{graphicx} % DO NOT CHANGE THIS
\usepackage{graphicx}  % DO NOT CHANGE THIS
\nocopyright
\usepackage{subfigure}
\usepackage{multirow}

\title{Dynamic Malware Analysis with Feature Engineering and Feature Learning}

\author{Zhaoqi Zhang, 
\Large \textbf{Panpan Qi}, 
\Large \textbf{Wei Wang}\\
School of Computing\\
National University of Singapore\\
zhaoqi.zhang@u.nus.edu, 
qipanpan@u.nus.edu, 
wangwei@comp.nus.edu.sg\\
}

\begin{document}
\maketitle
\begin{abstract}
Dynamic malware analysis executes the program in an isolated environment and monitors its run-time behaviour (e.g. system API calls) for malware detection. This technique has been proven to be effective against various code obfuscation techniques and newly released (``zero-day") malware. However, existing works typically only consider the API name while ignoring the arguments, or require complex feature engineering operations and expert knowledge to process the arguments. In this paper, we propose a novel and low-cost feature extraction approach, and an effective deep neural network architecture for accurate and fast malware detection. Specifically, the feature representation approach utilizes a feature hashing trick to encode the API call arguments associated with the API name. The deep neural network architecture applies multiple Gated-CNNs (convolutional neural networks) to transform the extracted features of each API call. The outputs are further processed through bidirectional LSTM (long-short term memory networks) to learn the sequential correlation among API calls. Experiments show that our solution outperforms baselines significantly on a large real dataset. Valuable insights about feature engineering and architecture design are derived from the ablation study. 
\end{abstract}

\section{Introduction}

Cybersecurity imposes substantial economic cost all over the world. A report~\cite{cea2018} from the United States government estimates that costs by malicious cyber activities in the U.S. economy lay between \$57 billion and \$109 billion in 2016. Malicious software (or malware) is one of the major cybersecurity threats that evolves rapidly. It is reported that more than 120 million new malware samples are being discovered every year~\cite{AV-TESTTheIndependendIT-SecurityInstitute2017}. Therefore, the development of malware detection techniques is urgent and necessary. 

Researchers have been working on malware detection for decades. The mainstream solutions include static analysis and dynamic analysis. Static analysis methods scan the binary byte-streams of the software to create signatures, such as printable strings, n-gram, instructions, etc~\cite{kruegel2005polymorphic}. However, the signature-based static analysis might be vulnerable to code obfuscation~\cite{rhode2018early,gibert2018classification} or inadequate to detect new (``zero-day") malware~\cite{vinod2009survey}. In contrast, dynamic analysis algorithms execute each software in an isolated environment (e.g., a sandbox) to collect its run-time behaviour information. By using behaviour information, dynamic analysis exerts a higher detection rate and is more robust than static analysis~\cite{damodaran2017comparison}. In this paper, we focus on dynamic analysis.

Among behaviour information, the system API call sequence is the most popular data source as it captures all the operations (including network access, file manipulation operations, etc.) executed by the software. Each API call in the sequence contains two important parts, the API name and the arguments. Each API may have zero or multiple arguments, each of which is represented as a name-value pair. To process behaviour information, a lot of feature engineering methods are proposed. For example, if we consider the API name as a string, then the most N (e.g., 1000) frequent n-gram features can be extracted ($n=1, 2, \cdots$) from the sequence. However, it is non-trivial to extract the features from the arguments of heterogeneous types, including strings, integers, addresses, etc. 

Recently, researchers have applied deep learning models to dynamic analysis. Deep learning models like convolutional neural network (CNN) and recurrent neural network (RNN) can learn features from the sequential data directly without feature engineering. Nonetheless, the data of traditional deep learning applications like computer vision and natural language processing is homogeneous, e.g., images (or text). It is still challenging to process the heterogeneous API arguments using deep learning models. Therefore, most existing approaches ignore the arguments. There are a few approaches~\cite{tian2010differentiating,fang2017new,agrawal2018neural} leveraging API arguments. However, these approaches either treat all arguments as strings~\cite{tian2010differentiating,agrawal2018neural} or only consider the statistical information of arguments~\cite{ahmed2009using,tian2010differentiating,islam2013classification}.
They consequently cannot fully exploit the heterogeneous information from different types of arguments.

In this paper, we propose a novel feature engineering method and a new deep learning architecture for malware detection. In particular, for different types of arguments, our feature engineering method leverages hashing approaches to extract the heterogeneous features separately. The features extracted from the API name, category, and the arguments, are further concatenated and fed into the deep learning model. We use multiple gated CNN models~\cite{dauphin2017language} to learn abstract lower dimensional features from the high dimensional hash features for each API call. The output from the gated CNN models is processed by a bidirectional LSTM to extract the sequential correlation of all API calls. 

Our solution outperforms all baselines with a large margin. Through extensive ablation study, we find that both feature engineering and model architecture design are crucial for achieving high generalization performance.

The main contributions of this paper include:
\begin{enumerate}
    \item We propose a novel feature representation for system API arguments. The extracted features from our dataset will be released for public access.
    \item We devise a deep neural network architecture to process the extracted features, which combines multiple gated CNNs and a bidirectional LSTM. It outperforms all existing solutions with a large margin.
    \item We conduct extensive experiments over a large real dataset\footnote{\url{https://github.com/joddiy/DynamicMalwareAnalysis} is the link of the code and the dataset.}. Valuable insights about the feature and model architecture are found through ablation study.
\end{enumerate}

\section{Related Work}

%[scale=0.69]
\begin{figure*}[h]
\centering
\includegraphics[width=1.8\columnwidth]{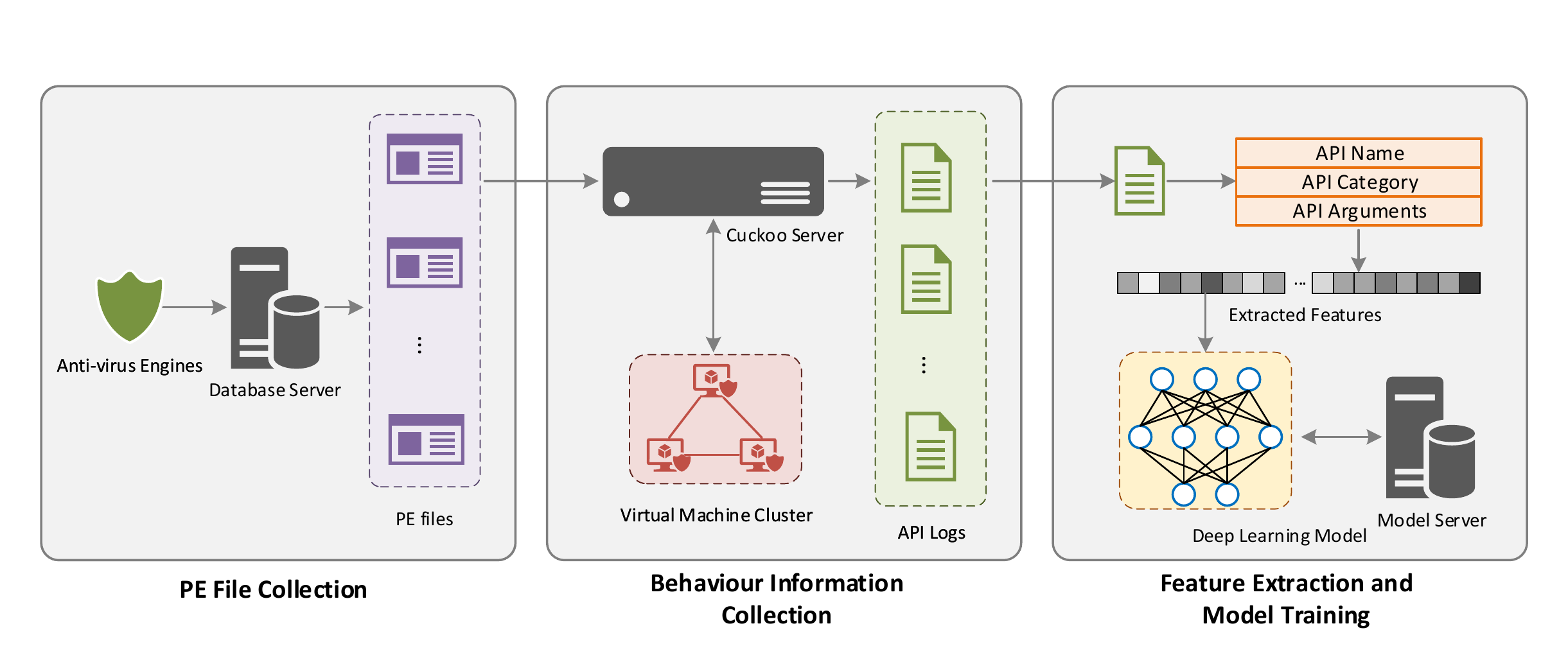}
\caption{System Architecture} 
\label{fig:framework_architecture}
\end{figure*}
In this section, we review the dynamic malware analysis from the feature engineering and the deep learning perspectives. 

\subsection{Feature Engineering for API Calls}

\cite{trinius2009malware} introduce a feature representation called Malware Instruction Set (MIST). MIST uses several levels of features to represent a system call. The first level represents the category and name of the API call. The following levels are specified manually for each API call to represent their arguments. However, for different APIs, features at the same level may indicate different types of information. The inconsistency imposes challenges to learn patterns using machine learning models.

\cite{qiao2013analyzing} extend the MIST and propose a representation called Byte-based Behaviour Instruction Set (BBIS). They claim that only the first level (the category and name of an API call) of MIST is effective. Besides, they propose an algorithm CARL to process consecutively repeated API calls.

Statistical features are popular for training machine learning models. Strings from the API call's name and its arguments are extracted to calculate the frequency and distribution as the features in~\cite{tian2010differentiating,islam2010classification,islam2013classification}.~\cite{ahmed2009using} also use statistical features that capture both the spatial and temporal information. Spatial information is extracted from arguments, such as the mean, variance, and entropy. Temporal information is from the n-gram API calls, including the correlation and transformation possibility between two n-gram API calls.

\cite{salehi2012miner} propose a feature representation associating the API calls with their arguments. It concatenates each argument with the name of its API call to form a new sequence, However, this approach leads to an extremely long feature vector and might lose the pattern of API call sequence. \cite{hansen2016approach} propose another two feature representations. These representations consist of first 200 API calls as well as its ``argument". However, this ``argument" only indicates whether this API call is connected with the later one, while ignoring the original arguments.

\subsection{Deep Learning Based Approaches}

\cite{david2015deepsign} treat the sandbox report as an entire text string, and then split all strings by any special character. They count the frequency of each string and use a 20,000-bit vector to represent the top 20,000 frequent ones. Their model is a deep belief network (DBN) which consists of eight layers (from 20,000-sized vectors to 30-sized vectors). Cross-entropy loss is used to train the model. They attain 98.6\% accuracy on a small dataset with 600 test samples.

\cite{pascanu2015malware} propose a two-stage approach, a feature learning stage and a classification stage. At the first stage, they use RNNs to predict the next possible API call based on the previous API call sequence. At the classification stage, they freeze the RNNs, and feed the outputs into a max-pooling layer to transform features for classification. They attain 71.71\% recall rate at a false positive rate of 0.1\% on a dataset with 75,000 samples.

\cite{kolosnjaji2016deep} propose an approach which combines CNN with LSTM. Their approach stacks two CNN layers, and each CNN layer uses a 3-sized kernel to simulate the 3-gram approach. After the CNN, an LSTM with a 100-sized hidden vector is appended to handle the time-series sequence. 

The previous papers typically ignore arguments.~\cite{huang2016mtnet} use a feature representation with three parts, the presence of runnable code in arguments, the combination of the API call name with one of its arguments (selected manually), and the 3-gram of API call sequence. This feature representation is reduced from 50,000 to 4,000 by a random projection. ~\cite{agrawal2018neural} propose a feature representation with a one-hot vector from API call name and top N frequent n-gram of the argument strings. The model uses several stacked LSTMs that shows a better performance than~\cite{kolosnjaji2016deep}. They also claim that multiple LSTMs cannot increase the performance.

\section{System Framework}

To collect the run-time API calls, we implement the system shown in  Figure~\ref{fig:framework_architecture}. The system has three parts, PE files collection, behaviour information collection, and feature extraction as well as model training. 

\subsection{PE Files Collection}\label{sec:pefile}

The workflow of our system starts from the portable executable (PE) files collection. In this paper, we focus on detecting malware in portable executable (PE) file format in Windows systems, which is the most popular malware file format~\cite{AV-TESTTheIndependendIT-SecurityInstitute2017}. This collection part has been implemented by a local anti-virus company, SecureAge Technology of Singapore. In addition, the company maintains a platform with 12 anti-virus engines to classify the PE files. The classification results are aggregated to get the label of each PE file for model training. Once the model is trained, it will be added into the platform as the 13th anti-virus engine. After the collection, an execution queue is maintained to submit the PE files for execution. It monitors the storage usage and decides whether to execute more PE files.

\subsection{Behaviour Information Collection}

Cuckoo\footnote{\url{https://cuckoosandbox.org/}}, an open-source software, is used to run the PE files and gather execution logs. It executes PE files inside virtual machines and uses API hooks to monitor the API call trace (i.e., the behaviour information). Besides, Cuckoo simulates some user actions, such as clicking a button, typing some texts, etc. In our system, we maintain dozens of virtual machines on each server. All virtual machines are installed with a 64-bit Windows 7 system and several daily-use software. We leverage the snapshot feature of the virtual machine to roll it back after execution. All generated logs are stored locally on the Cuckoo server.

\subsection{Feature Extraction and Model Training}

The execution logs generated by the sandbox contain detailed runtime information of the PE files, 
whose size ranges from several KB to hundred GB. We design a feature engineering solution that can run in parallel to extract features from the raw execution logs efficiently. Once the features are extracted, we train our deep learning model on a model server with GPUs for malware classification. 

\section{Methodology}

\subsection{Feature Engineering}\label{sec:feature}

Most previous works~\cite{qiao2013analyzing,pascanu2015malware,kolosnjaji2016deep} neglect the arguments of the API call, and only consider the API name and category. Consequently, some important (discriminative) information is lost~\cite{agrawal2018neural}. For example, the features of two write operations (API calls) would be exactly the same if the file path argument is ignored. However, the write operation might be benign when the target file is created by the program itself but be malicious if the target file is a system file.
A few works~\cite{trinius2009malware,agrawal2018neural,huang2016mtnet} that consider the arguments fail to exploit the heterogeneous information from different types of arguments. 

We propose to adapt the hash method from \cite{weinberger2009feature} to encode the name, category and arguments of an API separately. As shown in Table \ref{table:feature}, our feature representation consists of different types of information. The API name has 8 bins, and the API category has 4 bins. The API arguments part has 90 bins, 16 for the integer arguments and 74 for the string arguments. For the string arguments, several specific types of strings (file path, Dlls, etc.) are processed. Besides, 10 statistical features are extracted from all printable strings. All these features are concatenated to form a 102-dimension feature vector.

\begin{table}[]
\caption{Feature representation overview}
\centering
\resizebox{0.85\columnwidth}!{
\begin{tabular}{|c|c|c|l|c|c|}
\hline
\multicolumn{4}{|c|}{\textbf{Feature Type}} & \textbf{Details} & \textbf{Dim} \\ \hline
API name & \multicolumn{3}{c|}{Strings} & \begin{tabular}[c]{@{}c@{}}Internal words \\ hashing trick\end{tabular} & 8 \\ \hline
API category & \multicolumn{3}{c|}{Strings} & Hashing trick & 4 \\ \hline
\multirow{7}{*}{\begin{tabular}[c]{@{}c@{}}API \\ Arguments\end{tabular}} & \multicolumn{3}{c|}{Integers} & Hashing trick & 16 \\ \cline{2-6} 
 & \multirow{6}{*}{Strings} & \multicolumn{2}{c|}{Paths} & \multirow{5}{*}{\begin{tabular}[c]{@{}c@{}}Hashing trick\\  with hierarchy\end{tabular}} & 16 \\ \cline{3-4} \cline{6-6} 
 &  & \multicolumn{2}{c|}{Dlls} &  & 8 \\ \cline{3-4} \cline{6-6} 
 &  & \multicolumn{2}{c|}{\begin{tabular}[c]{@{}c@{}}Registry\\ keys\end{tabular}} &  & 12 \\ \cline{3-4} \cline{6-6} 
 &  & \multicolumn{2}{c|}{Urls} &  & 16 \\ \cline{3-4} \cline{6-6} 
 &  & \multicolumn{2}{c|}{IPs} &  & 12 \\ \cline{3-6} 
 &  & \multicolumn{2}{c|}{\begin{tabular}[c]{@{}c@{}}String \\ statistics\end{tabular}} & \begin{tabular}[c]{@{}c@{}}numStrings, avLength,\\ numChars, entropy,\\ numPaths, numDlls,\\ numUrls, numIPs,\\ numRegistryKeys,\\ numMZ\end{tabular} & 10 \\ \hline
\end{tabular}
}
\label{table:feature}
\end{table}

\subsubsection{API Name and Category}

Cuckoo sandbox tracks 312 API calls in total which belong to 17 categories. Each API name consists of multiple words with the first letter of each word capitalized, such as ``GetFileSize". We split the API name into words and then process these words by applying the feature hashing trick below. For the API category, since the category typically is a single word, for example, ``network", we split the word into characters and apply the feature hashing trick. In addition, we compute the MD5 value of the API name, category and arguments to remove any consecutively repeated API calls.

We use feature hashing~\cite{weinberger2009feature} in Equation \ref{eq:1} to encode a sequence of strings into a fixed-length vector. The random variable $x$ denotes a sequence of elements, where each element is either a string or a character. $M$ denotes the number of bins, i.e., 8 for API name, and 4 for API category. The value of the $i$-th bin is calculated by:

\begin{equation}\label{eq:1}\phi_i(x)=\sum_{j:h(x_j)=i}\xi(x_j)\end{equation}

where $h$ is a hash function that maps an element, e.g., $x_j$, to a natural number $m \in \{1, ..., M\}$ as the bin index; $\xi$ is another hash function that maps an element to $\{\pm 1\}$. That is, for each element $x_j$ of $x$ whose bin index $h(x_j)$ is $i$, we add $\xi(x_j)$ into the bin.

\subsubsection{API Arguments}

As for API arguments, there are only two types of values, namely integers and strings. The individual value of an integer is meaningless. The argument name is required to get the meaning of the value. The same integer value might indicate totally different semantics with different argument names. For example, number 22 with the name ``port" is different from the one with the name ``size". 

We adapt the previous feature hashing method to encode the integer's argument name as well as its value, as shown in Equation \ref{eq:2}. We use the argument name to locate the hash bin. In particular, we use all the arguments whose names' hash value is $i$ to update the i-th bin via summation. For each such argument, we compute the contribution to the bin as shown in Equation~\ref{eq:2}, where $\xi(x_j^{name})$ is a hash function over the argument name and $x_j^{value}$ is the value of the integer argument. Because integers may distribute sparsely within a range, we normalize the value using the logarithm to squash the range.

\begin{equation}\label{eq:2}\phi_i(x)=\sum_{j:h(x_j^{name})=i}\xi(x_j^{name})\log(|x_j^{value}|+1)\end{equation} where $h$ and $\xi$ are the same hash functions as in Equation \ref{eq:1}.

For strings of API arguments, their values are more complicated than integers. Some strings starting with `0x' contain the address of some objects. And some other may contain the file path, IP address, URL, or plain text. Besides, some API arguments may even contain the content of an entire file. The variety of strings makes it challenging to process them. According to the previous work~\cite{tian2010differentiating,islam2010classification,islam2013classification,ahmed2009using}, the most important strings are the values about file paths, DLLs, registry keys, URLs, and IP addresses. Therefore, we use the feature hashing method in Equation \ref{eq:1} to extract features for these strings.

To capture the hierarchical information contained in the strings, we parse the whole string into several substrings and process them individually. For example, we use ``\url{C:\\}" to identify a file path. For a path like ``\url{C:\\a\\b\\c}", four substrings are generated, namely ``\url{C:}", ``\url{C:\\a}", ``\url{C:\\a\\b}", and ``\url{C:\\a\\b\\c}". All these substrings are processing by Equation \ref{eq:1}. The same processing method is applied for DLLs, registry keys and IPs. The DLLs are strings ending with ``.dll". The registry keys often start with ``HKEY\_". IPs are those strings with four numbers (range from 0 to 255) separated by dots. Slightly different for URLs, we only generate substrings from the hostname of the URL. For example, for ``\url{https://security.ai.cs.org/}", the following substrings will be generated ``org", ``\url{cs.org}", ``\url{ai.cs.org}" and ``\url{security.ai.cs.org}". In this way, the domain and organization information will contribute more to the feature.

For lots of other types of strings, based on the previous work\cite{ahmed2009using,tian2010differentiating,islam2010classification}, we extract statistical information from all the printable strings. The printable strings consist of characters ranging from 0x20 to 0x7f. Therefore, all the paths, registry keys, URLs, IPs and some other printable strings are included. One type of strings starting with ``MZ" is often a buffer that contains an entire PE file and usually occurs in malicious PE files such as thread injection\cite{liu2011behavior}. Therefore, we additionally count the occurrences of ``MZ" strings. A 10-dimension vector is used to record the number of strings, their average length, the number of characters, the entropy of characters across all printable strings, and the number of paths, DLLs, URLs, registry keys, IPs and ``MZ" strings. 

%The semantic information of virtual address is hard to capture. Structs,  has no
We have not handled other arguments such as virtual addresses, structs, et al., which are relatively not so important compared with above types of arguments. Although the proposed feature engineering method is easy to be applied to them using extra bins, we look forward to more targeted researches to explore these arguments.

\subsection{Model Architecture}\label{sec:model}
%[scale=0.8]
\begin{figure}[]
\centering
\includegraphics[width=0.7\columnwidth]{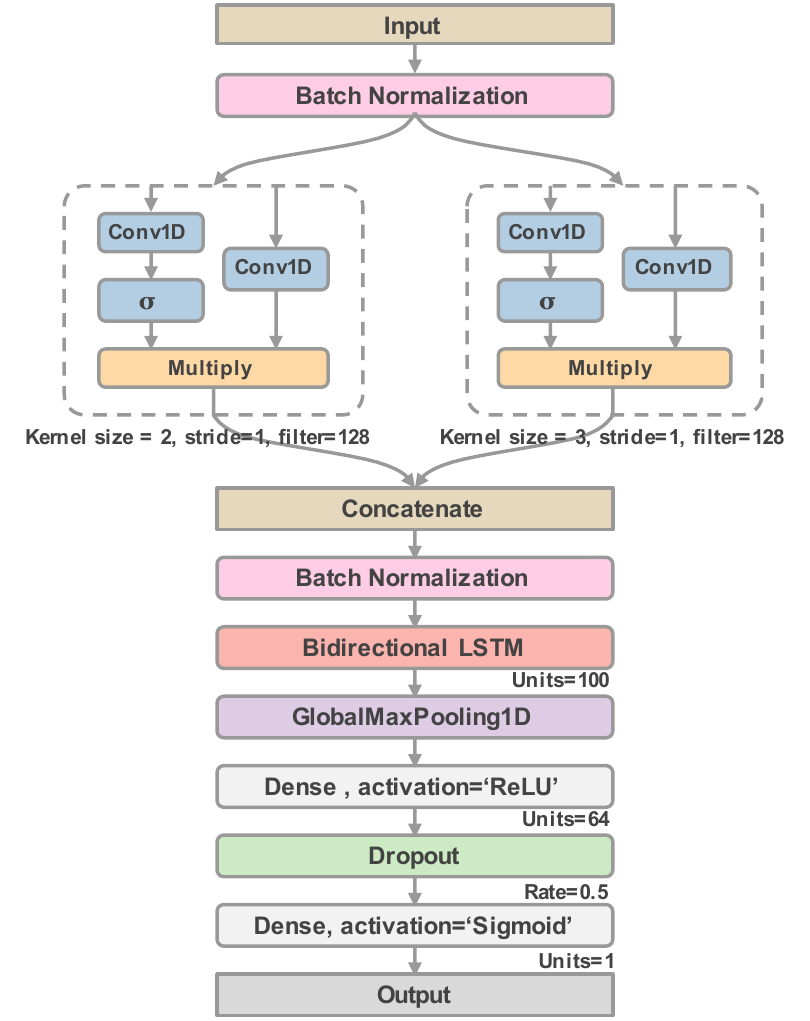}
\caption{An illustration of the proposed model}
\label{fig:modeloverview}
\end{figure}

We present a deep neural network architecture that leverages the features from the proposed feature engineering step. Figure~\ref{fig:modeloverview} is an overview of our proposed deep learning model. 

\subsubsection{Input Module}

After feature engineering, we get the input vector whose size is $(N, d)$, where $N$ is the length of the API call sequence, and $d$ (102 bits) is the dimension of each extracted API feature. We first normalize the input by a batch normalization layer~\cite{ioffe2015batch}. This batch normalization layer normalizes the input values by subtracting the batch mean and dividing by the batch standard deviation. It makes sure some dimensions of the feature vector are not so large to affect the training; it also has a regularization effect, which is validated in the experiments. 

\subsubsection{Gated-CNNs Module}

Several gated-CNNs~\cite{dauphin2017language} are applied after the input module. Gated-CNNs allows the selection of important and relevant information making it competitive with recurrent models on language tasks but consuming less resource and less time.

For each gated CNN, the input is fed into two convolution layers respectively. Let $X_A$ denotes the output of the first convolution layer, and $X_B$ denotes the output of the second one; they are combined by $X_A \otimes \sigma(X_B)$, which involves an element-wise multiplication operation. Here, $\sigma$ is the sigmoid function $\sigma(x)=\frac{1}{1+e^{-x}}$. $\sigma(X_B)$ is regarded as the gate that controls the information from $X_A$ passed to the next layer in the model. 

Following the idea in~\cite{shen2014latent}, 1-D convolutional filters are used as n-gram detectors. As Figure~\ref{fig:modeloverview}, we use two gated CNNs whose filter size is 2 and 3 respectively. All convolution layers' filter size is 128, and stride is 1. 

\subsubsection{Bi-LSTM Module}

All outputs from Gate CNNs are concatenated together. A batch normalization layer is applied to these outputs to reduce overfitting. We use bidirectional LSTM to learning sequential patterns. The number of units of each LSTM is 100.

LSTM is a recurrent neural network architecture, in which several gates are designed to control the information transmission status so that it is able to capture the long-term context information~\cite{pichotta2016learning}. Bidirectional LSTM is two LSTMs stacking together but with different directional input. Compared to unidirectional LSTM, bidirectional LSTM is able to integrate the information from past and future states simultaneously. Bidirectional LSTM has been proved effective at malware detection by~\cite{agrawal2018neural}.

\subsubsection{Classification Module}

After learning sequential patterns from Bi-LSTM module, a global max-pooling layer is applied to extract abstract features from the hidden vectors. Instead of using the final activation of the Bi-LSTM, a global max-pooling layer relies on each signal observed throughout the sequence, which helps retain the relevant information learned throughout the sequence.

After the global max-pooling layer, we use a dense layer with units number 64 to reduce the dimension of the intermediate vector to 64. A ReLU activation is applied to this dense layer. Then we use a dropout layer with a rate of 0.5 to reduce overfitting. Finally, a dense layer with units number 1 reduces the dimension to 1. A Sigmoid activation is appended after the dense layer to output the probability. 

Our model is supervised with the label associated with each input vector. To measure the loss for training the model, binary cross-entropy function is used as Equation \ref{eq:3}.

\begin{equation}\label{eq:3}\ell(X, y)=-(y \log(P[Y=1|X])+(1-y)log(P[Y=0|X]))\end{equation}

In addition, the optimization method we take is Adam, and the learning rate is 0.001. 

\section{Experiments}

\subsection{Dataset}

As described before, 12 commercial anti-virus engines are set up to classify the PE file. We set a PE file as positive if 4 or more engines agree that it is malicious. And if none of the engines classifies it as malware, we set it as negative. For other cases, we think the results are inconclusive and therefore exclude them from our dataset. 

\begin{table}[h]
\centering
\caption{Summary of the data}
\resizebox{0.6\columnwidth}!{\begin{tabular}{|c|c|c|}
\hline
\textbf{Dataset} & \textbf{Positive files} & \textbf{Negative files} \\ \hline
April & 15931 & 11417 \\ \hline
May & 11856 & 21983 \\ \hline
\end{tabular}}
\label{table:1}
\end{table}

The collected data are archived by the date and we pick two months (April and May) data to conduct our experiments. All these PE files are processed by our system (as shown in Figure~\ref{fig:framework_architecture}) to collect the API call sequences. Table \ref{table:1} is a summary of the data, where the row represents the statistics of the data in a month.

\subsection{Model Evaluation}

\begin{table*}[h]
\centering
\caption{The experimental results}
\label{table:experimental_results}
\resizebox{1.9\columnwidth}!{
\begin{tabular}{|c|c|c|c|c|c|c|c|c|c|}
\hline
\multirow{2}{*}{\textbf{Type}}             & \multirow{2}{*}{\textbf{Approach}} & \multirow{2}{*}{\textbf{Arguments}} & \multicolumn{3}{c|}{\textbf{4-fold CV Performance}} & \multicolumn{3}{c|}{\textbf{Test Performance}} & \multirow{2}{*}{\begin{tabular}[c]{@{}c@{}}Inference Time\\ (ms/sample)\end{tabular}}\\ \cline{4-9}
                                  &                           &                            & AUC(\%)          & ACC(\%)          & Recall(\%)       & AUC(\%)         & ACC(\%)        & Recall(\%)     &                                                                                       \\ \hline
\multirow{3}{*}{\begin{tabular}[c]{@{}c@{}}Machine\\ Learning\end{tabular}} & \cite{uppal2014malware}     & No                         & 96.18$\pm$0.26    & 90.89$\pm$0.57    & 1.37$\pm$0.39     & 94.02$\pm$0.62   & 86.01$\pm$0.31  & 1.61$\pm$0.74   & 98.56                                                                              \\ \cline{2-10} 
                                  & \cite{tian2010differentiating}    & \multirow{2}{*}{Yes}       & 99.10$\pm$0.07    & 95.83$\pm$0.29    & 82.42$\pm$3.12    & 97.77$\pm$0.07   & 93.18$\pm$0.17  & 67.45$\pm$2.38  & 123.03                                                                              \\ \cline{2-2} \cline{4-10} 
                                  & \cite{fang2017new}      &                            & 98.62$\pm$0.06    & 94.58$\pm$0.55    & 53.51$\pm$26.28    & 97.02$\pm$0.11   & 90.88$\pm$0.28  & 41.72$\pm$1.76  & 116.01                                                                               \\ \hline
\multirow{4}{*}{\begin{tabular}[c]{@{}c@{}}Deep\\ Learning\end{tabular}}    & \cite{pascanu2015malware}    & \multirow{2}{*}{No}        & 95.35$\pm$1.65    & 89.06$\pm$1.80    & 9.16$\pm$15.61     & 50.69$\pm$24.64   & 32.21$\pm$6.40  & 0.67$\pm$0.61   & 94.23                                                                                \\ \cline{2-2} \cline{4-10} 
                                  & \cite{kolosnjaji2016deep}   &                            & 98.82$\pm$0.15    & 95.34$\pm$0.74    & 59.48$\pm$18.86    & 97.58$\pm$1.46   & 93.32$\pm$2.38  & 42.49$\pm$16.04  & 92.19                                                                                \\ \cline{2-10} 
                                  & \cite{agrawal2018neural}    & \multirow{2}{*}{Yes}       & 99.07$\pm$0.13    & 95.87$\pm$0.28    & 77.78$\pm$9.55    & 98.19$\pm$0.32   & 94.86$\pm$0.22  & 60.11$\pm$6.69  & 257.57                                                                               \\ \cline{2-2} \cline{4-10} 
                                  & Proposed Model            &                            & \textbf{99.46$\pm$0.04}    & \textbf{96.76$\pm$0.26}    & \textbf{88.75$\pm$2.65}    & \textbf{98.71$\pm$0.17}   & \textbf{95.33$\pm$0.40}  & \textbf{71.48$\pm$3.08}  & 129.21                                                                               \\ \hline
\end{tabular}}
\end{table*}

%[scale=0.382]
\begin{figure*}[h]
\centering
\subfigure[Validation ROC curve]{
\label{fig:auc:val}
\includegraphics[width=0.8\columnwidth]{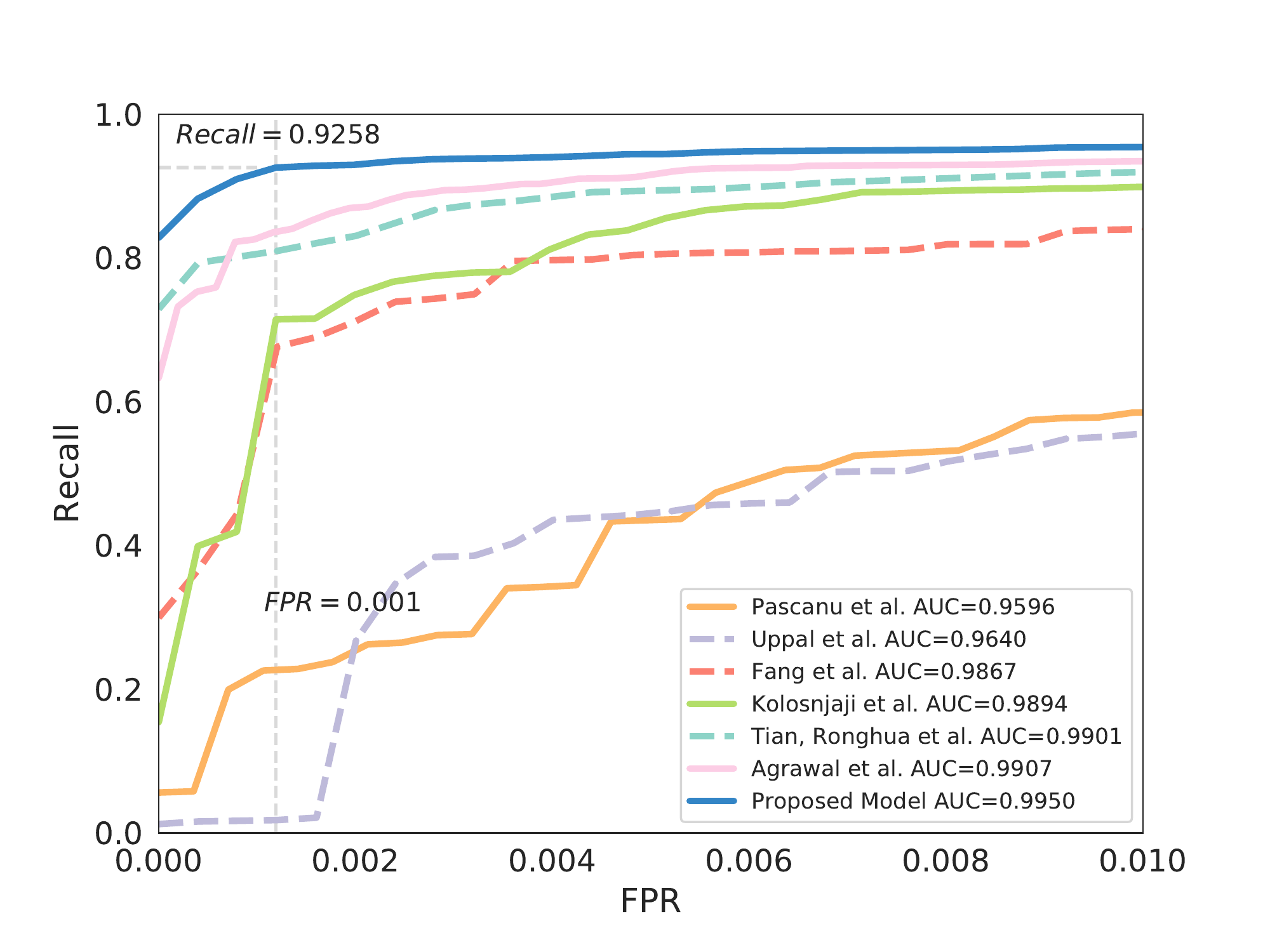}}
\subfigure[Test ROC curve]{
\label{fig:auc:test}
\includegraphics[width=0.8\columnwidth]{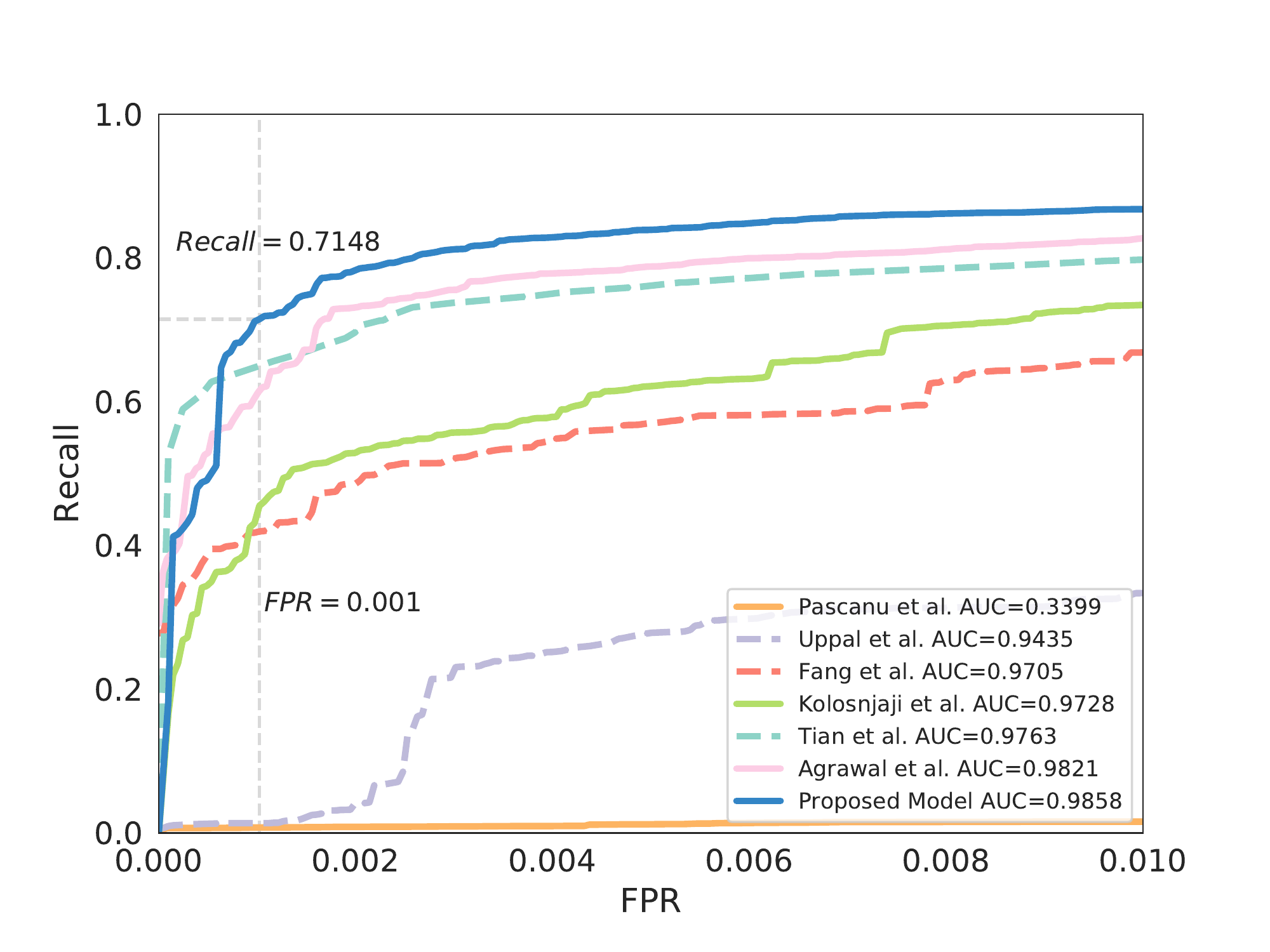}}
\caption{Comparisons of ROC curve of different models}

\label{fig:auc}
\end{figure*}

In order to investigate the performance improvement, we compare the proposed model with three machine learning-based models and three deep learning-based models. 

\begin{itemize}
    \item \textbf{\cite{uppal2014malware}} extract 3-gram vectors from API call names. Then they use the odds ration to select the most important vectors. SVM is applied as the model.
    \item \textbf{\cite{tian2010differentiating}} use a hash table to indicate the presence of strings. The strings come from both API names and arguments. The generated hash table is then used as features and the classifier is Random Forest.
    \item \textbf{\cite{fang2017new}} use hashing trick to map the API call names, return value and module name (a part of the arguments) into some fixed-size bins. Then top important features are selected and fed into XGBoost.
    \item \textbf{\cite{pascanu2015malware}} train a language model using RNN which can predict the next API call given the previous API calls. Then the RNN model is freezed and the hidden features are extracted for malware detection. The input of the model is a sequence of $d$-dimensional one-hot vectors whose elements are all zeros except the position (the element value is 1) for the corresponding API call.
    \item \textbf{\cite{kolosnjaji2016deep}} propose a model which combines stacked CNNs and RNNs. The input is also one-hot vectors for the API call sequence.
    \item \textbf{\cite{agrawal2018neural}} extract one-hot vectors from the API call sequence and frequent n-gram vectors from the API arguments. The model uses several stacked LSTMs.
\end{itemize}

All the experiments are conducted against our dataset. We use 4-fold cross-validation (or CV) over the April dataset to train the models and do the testing over the May dataset. Considering that new malware is being generated over time, there could be many PE files for new malware in the May dataset. Therefore, the performance indicates the model's capability for detecting unknown malware in a certain degree. 

Three metrics are considered: ROC (receiver operating characteristic curve) AUC (Area Under the Curve) score, ACC (accuracy) and Recall when FP (false positive) rate is 0.1\%. The recall is defined as the ratio of the correctly detected malware PE files over all malware PE files. The FP rate is the ratio of benign PE files incorrectly identified as malware. Anti-virus products are required to keep a low false alarm rate to avoid disturbing users frequently~\cite{Charles2017Malware}. A good model should achieve a high recall rate for a fixed low false positive rate. We provide 95\% confidence intervals for all these three metrics. In addition, the inference time per sample, which includes the time for feature processing and model prediction, is also taken into account.

From the experimental results in Table~\ref{table:experimental_results}, our proposed model achieves the best AUC score, accuracy and recall among all the baseline models at both CV and test dataset. 

Figure~\ref{fig:auc} displays the ROC curve of all models. The dashed curves are the ROCs of those traditional machine learning models, while the solid lines are the ROCs of those deep learning models. The experimental results illustrate that the traditional machine learning approaches and deep learning approaches are comparable.  It should be noted that the model~\cite{tian2010differentiating} achieves quite good results by using a basic method to extract the string information. This indicates the importance of strings in feature processing. Therefore, we spend a lot of effort on the feature engineering of string data. The results also show that models with argument features generally outperform the ones neglecting arguments. The argument features increase the test AUC score of the traditional machine learning method by 3\% and also increased the test AUC score of deep learning by about 1\%. Therefore, including API arguments is necessary. 

Figure~\ref{fig:auc} shows a margin between the results on validation and test dataset. Since the training dataset is collected before the testing dataset so the test data is likely to include new malware PE file. However, our proposed solution achieves the best performance on the test dataset, which confirms the ability in detecting new and constantly evolving malware.

As for the inference time, models with the argument features take a slightly longer time. However, hundreds of milliseconds inference time are relatively small and acceptable, because the data collection using Cuckoo sandbox is time-consuming, and costs 3-5 minutes per sample. 
The training takes about 10 minutes per epoch, which could be easily reduced via distributed training~\cite{ooi2015singa}.

\subsection{Ablation Study}

The proposed model consists of several components that can be flexibly adjusted, e.g., the Gated CNNs, Bi-LSTM and Batch Normalization. In order to explore the effects of different configurations, we employ several sets of comparison experiments by fixing other structures and only changing the testing component. These results of these experiments serve as the basis for the decision of our final model structure. 

\begin{itemize}
    \item \textbf{Gated CNNs} with three sets experiments, the Gated CNNs only with kernel size 2 (2-GatedCNN), two Gated CNNs with kernel size 2 and 3 (2,3-GatedCNN), three Gated CNNs with kernel size 2, 3 and 4 (2,3,4-GatedCNN). 
    \item \textbf{Batch Normalization} with four sets experiments, the model without any batch normalization (BN) layer, without the first BN layer (after the input), without the second BN layer (after the Gated CNNs), and with both BN layers.
    \item \textbf{Bi-LSTM} with three sets experiments, the model with none Bi-LSTM, with one Bi-LSTM, and with two Bi-LSTM stacked.
\end{itemize} 

\begin{figure}[h]
\centering
\subfigure[Validation AUC]{
\label{fig:auc:ngram:val}
\includegraphics[width=0.46\columnwidth]{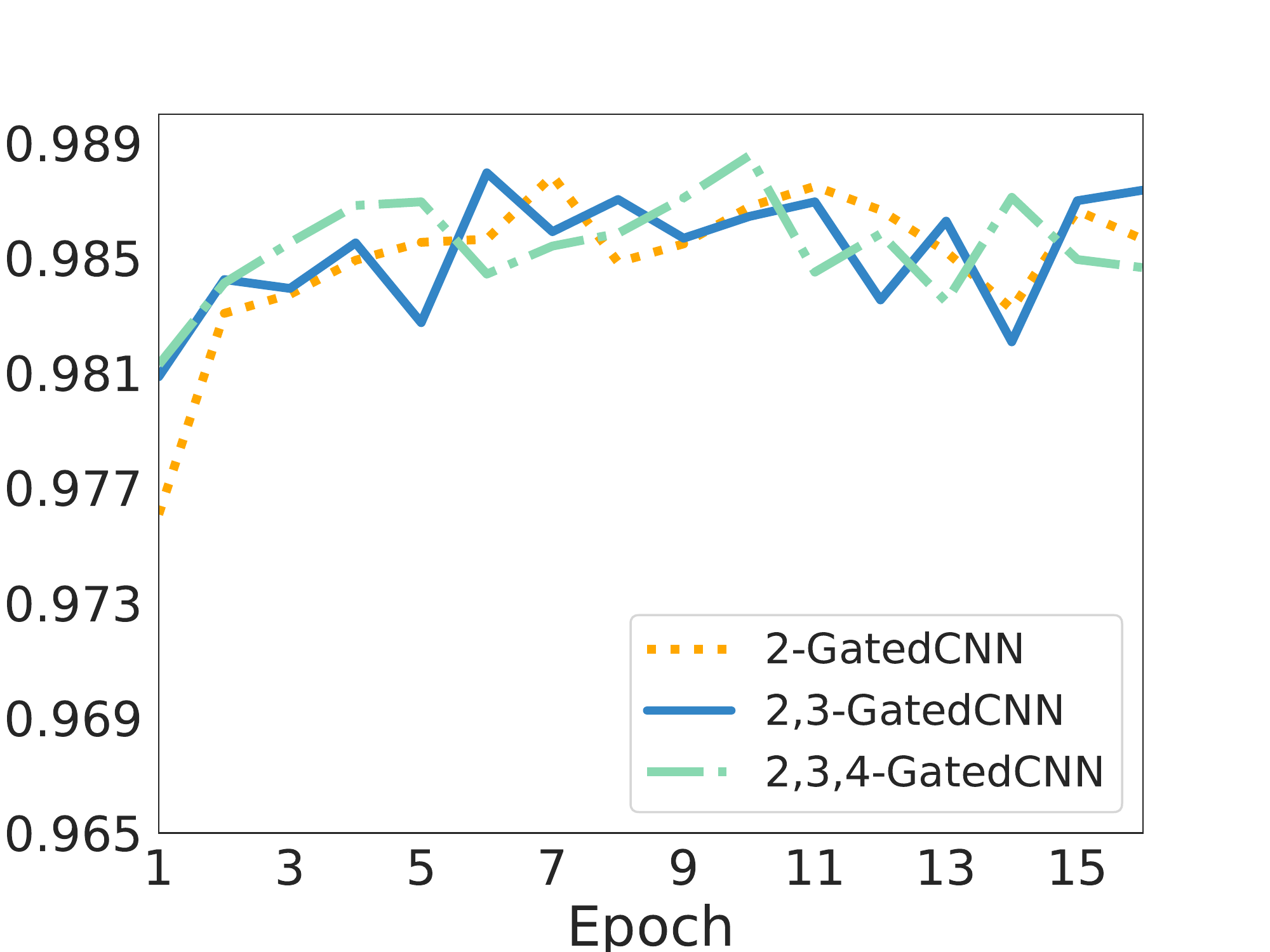}}
\subfigure[Test AUC]{
\label{fig:auc:ngram:test}
\includegraphics[width=0.46\columnwidth]{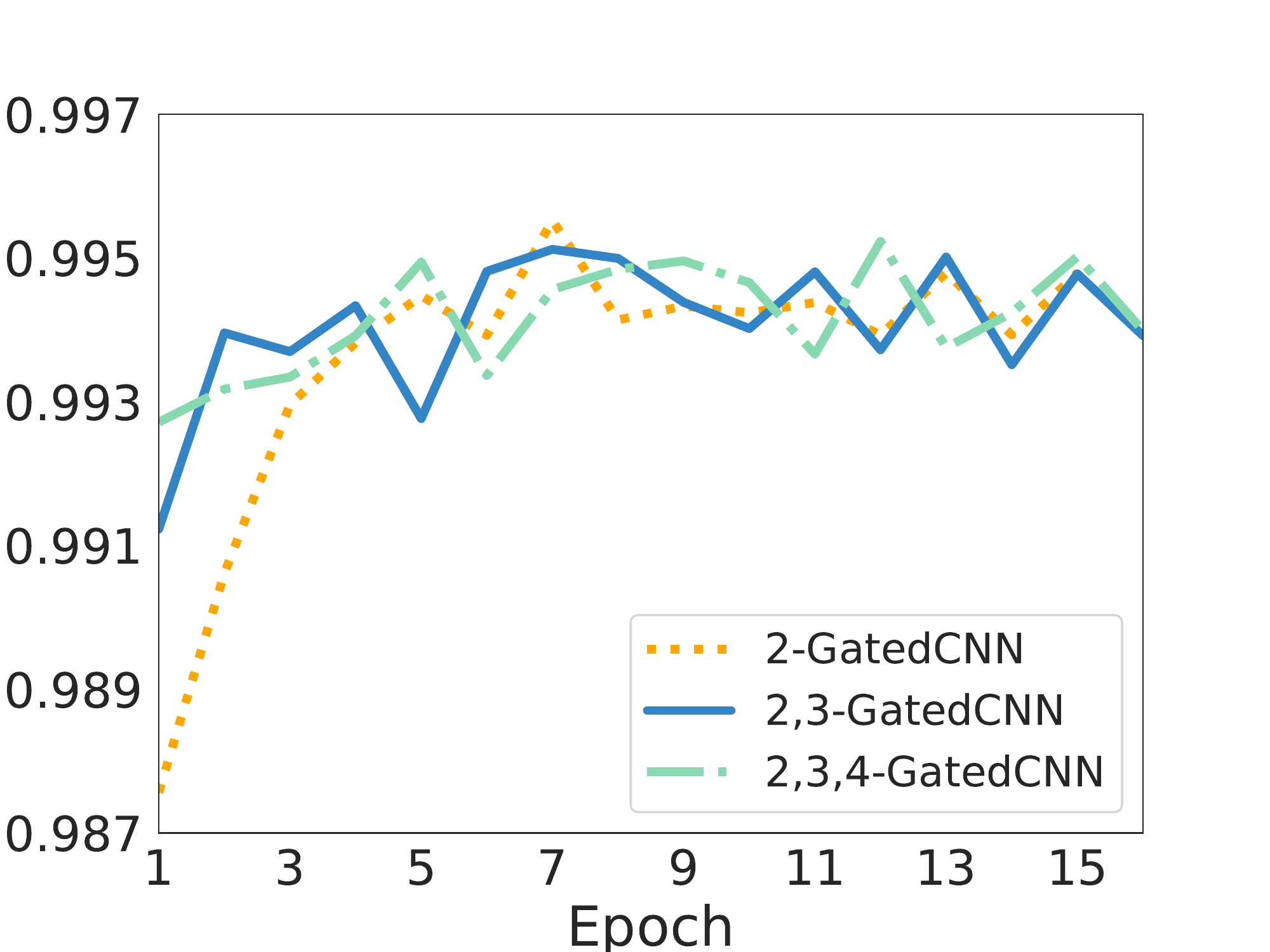}}
\caption{Comparison of AUC for Gated CNNs}

\label{fig:auc_ngram}
\end{figure}

\begin{figure}[h]
\centering
\subfigure[Validation AUC]{
\label{fig:auc:ngram:val}
\includegraphics[width=0.46\columnwidth]{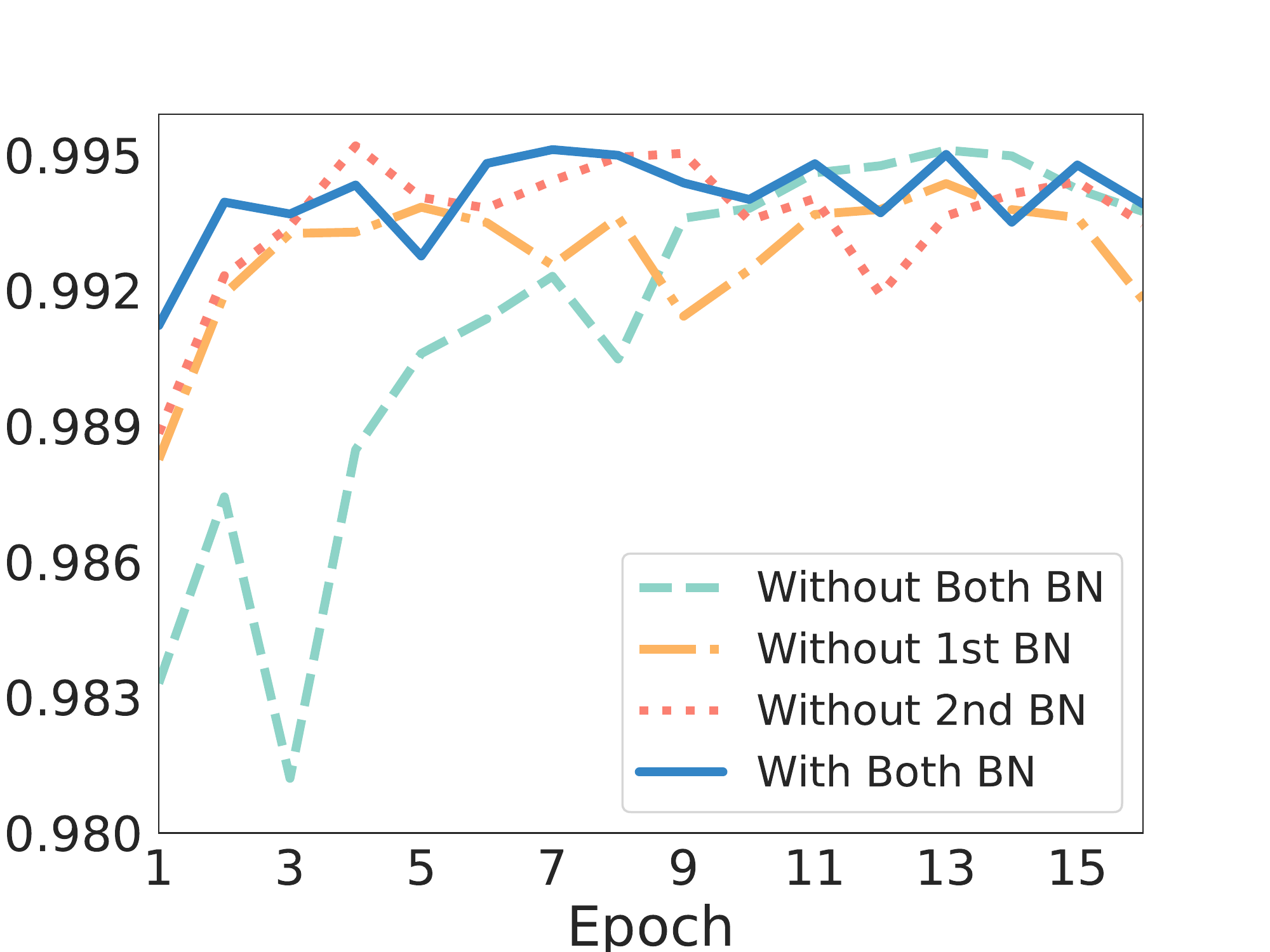}}
\subfigure[Test AUC]{
\label{fig:auc:ngram:test}
\includegraphics[width=0.46\columnwidth]{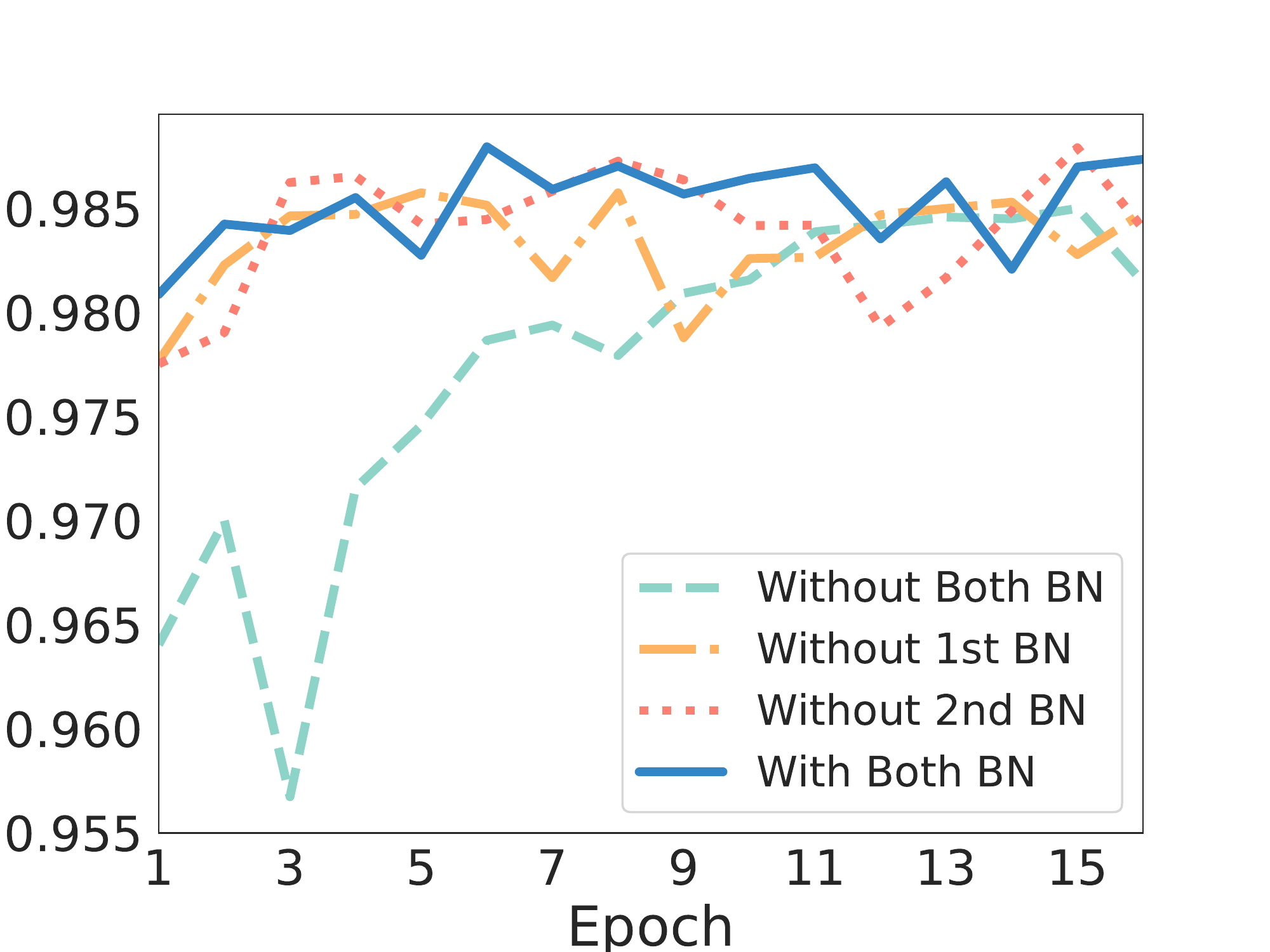}}
\caption{Comparison of AUC for Batch Normalization}

\label{fig:auc_norm}
\end{figure}

\begin{figure}[h]
\centering
\subfigure[Validation AUC]{
\label{fig:auc:ngram:val}
\includegraphics[width=0.46\columnwidth]{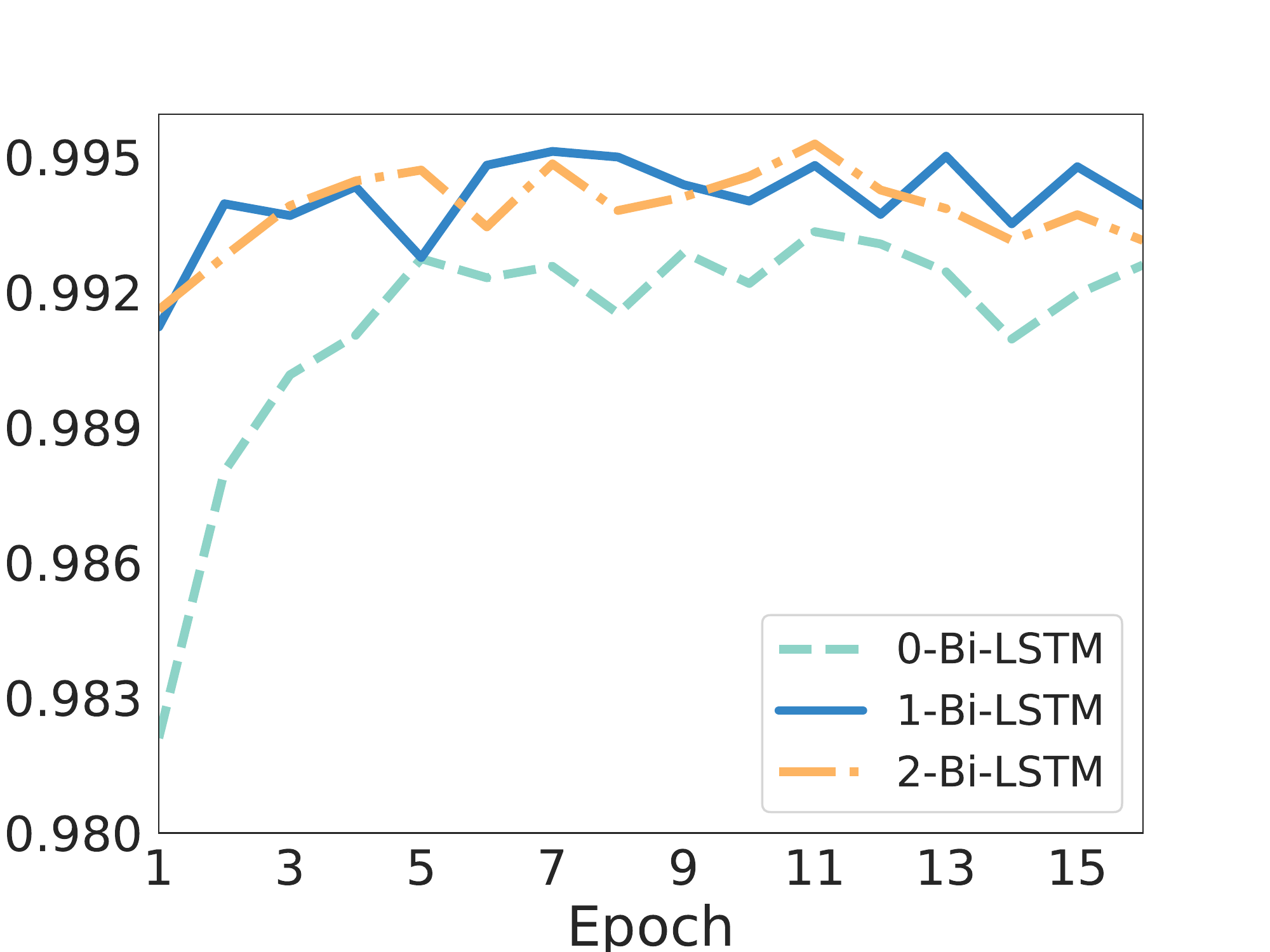}}
\subfigure[Test AUC]{
\label{fig:auc:ngram:test}
\includegraphics[width=0.46\columnwidth]{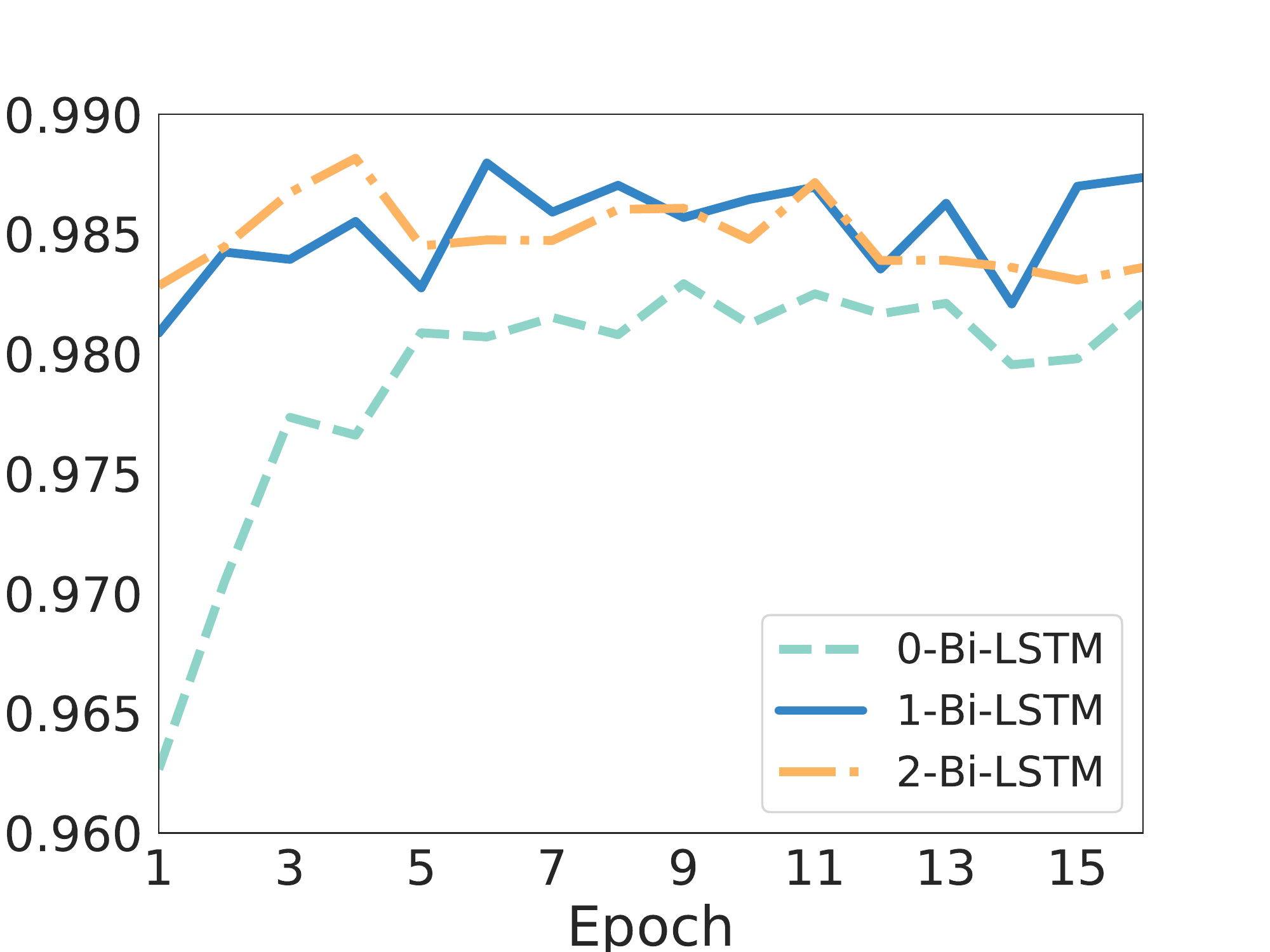}}
\caption{Comparison of AUC for Bi-LSTM}

\label{fig:auc_lstm}
\end{figure}

Figure~\ref{fig:auc_ngram} depicts the comparisons for different numbers of Gated CNNs. 2-GatedCNN converges slower although the final performance is very close to the other two models. In addition, increasing the number of gated CNN from 2,3-GatedCNN to 2,3,4-GatedCNN does not bring any performance improvement. The best AUC score of 2-GatedCNN and 2,3-GatedCNN is 98.80\% and 98.86\% respectively. Therefore, we choose 2,3-GatedCNN in our model.

Figure~\ref{fig:auc_norm} displays the performance with different numbers of batch normalization layers. Although these four curves tend to be closer at later epochs, the curve with both BN layers shows slightly superior performance with the highest AUC score at 98.80\%.

As for various numbers of Bi-LSTM, Figure~\ref{fig:auc_lstm} shows the performance for each configuration. Obviously, in both figures, the curve of 0-Bi-LSTM is below the other two curves by a large margin, which indicates the Bi-LSTM is vital. The other two curves in both figures are continuously staggered, however, 1-Bi-LSTM is slightly better with the highest point reaching 98.80\%. In addition, the computation time of 1-Bi-LSTM is 2 times faster than 2-Bi-LSTM. Thus, we choose 1-Bi-LSTM as the final configuration of the proposed model.

\section{Conclusion}

In this work, we propose a novel feature engineering method and a new deep learning architecture for malware detection over the API call sequence. Hashing tricks are applied to process the heterogeneous information from API calls, including the name, category and arguments. A homogeneous and low-cost feature representation is extracted. Then, we use multiple gated-CNNs to transform the high dimensional hash features from each API call, and feed the results into a Bi-LSTM to capture the sequential correlations of API calls within the sequence. The experiments show that our approach outperforms all baselines. Ablation study over multiple architecture variations verify our architecture design decisions.

\section*{Acknowledgements}

This work is supported by the National Research Foundation, Prime Ministers Office, Singapore under its National Cybersecurity RD Programme (No. NRF2016NCR-NCR002-020), and FY2017 SUG Grant. We also thank SecureAge Technology of Singapore for sharing the data.

\bibliographystyle{aaai}
\bibliography{4913_main}

\end{document}